\documentclass[11pt]{article}
\usepackage{amsmath,amssymb}
\usepackage{amsthm}
\usepackage{ascmac}
\usepackage{float}
\usepackage[pdftex]{graphicx}
\usepackage[usenames]{color}
\usepackage[color]{showkeys}
\usepackage[all]{xy}
\usepackage[pdftex,bookmarksnumbered=true,colorlinks=true,filecolor=blue,linkcolor=blue,urlcolor=blue,citecolor=blue,linktocpage=true]{hyperref}
\definecolor{refkey}{rgb}{1,1,1} %color for showkey

\numberwithin{equation}{section}

\newcommand{\la}{\lambda} %lambda

\newcommand{\Acal}{\mathcal{A}}

\newcommand{\Ocal}{\mathcal{O}}

\begin{document}

\definecolor{labelkey}{rgb}{1,1,1} %%これをコメントアウトすると、ref{}の名前が見えるようになる

\renewcommand{\thefootnote}{\fnsymbol{footnote}}
\setcounter{page}{0}
%%%%%%%%%%%%%%%%% Title page %%%%%%%%%%%%%%%%%%%%%%%%%%%%%%%%%
\thispagestyle{empty}
\begin{flushright} %paper number
RIKEN-iTHEMS-Report-19 \\
OU-HET-1009
\end{flushright}

\vskip3cm
\begin{center}
{\LARGE {\bf Comments on holographic entanglements \\ in cutoff AdS}}  %title
\vskip1.5cm
{\large %name address
\bf {Toshihiro Ota$^{a, b}$\footnote{\href{mailto:toshihiro.ota@riken.jp}{toshihiro.ota@riken.jp}}
}

\vskip1cm
\it ${}^{a}$RIKEN Interdisciplinary Theoretical \& Mathematical Sciences Program (iTHEMS), \\
 Wako, Saitama 351-0198, Japan \\ %affiliation
\it ${}^{b}$Department of Physics, Osaka University,  Toyonaka, Osaka 560-0043, Japan}
\end{center}

\vskip1cm
\begin{abstract} %abstract
We study two-interval holographic entanglement entropy and entanglement wedge cross section in cutoff AdS.
In particular, we investigate phase transitions of them.
For two-interval entanglement entropy, the transition point monotonically decreases with a deformation parameter, which means that by the $T\bar{T}$ deformation the degrees of freedom in subsystems are decreasing.
This implies that the effect of the $T\bar{T}$ deformation can be regarded as the rescaling of the energy scale.
We also study entanglement wedge cross section in cutoff AdS, and our result implies that for the entanglement of purification in the $T\bar{T}$ deformed CFTs phase transition could occur even for fixed subsystems.
\end{abstract}

%%%%%%%%%%%%%%%%%%%%%%%%%%%%%%%%%%%%%%%%%%%%%%%%%%%%%%%%
\renewcommand{\thefootnote}{\arabic{footnote}}
\setcounter{footnote}{0}

\vfill\eject

\tableofcontents

%%%%%%%%%%%%%%%%%% section 1 %%%%%%%%%%%%%%%%%%%%%%%%%%%%%%%%%
\section{Introduction}
\label{sec:intro}

The AdS/CFT correspondence \cite{Maldacena:1997re,Gubser:1998bc,Witten:1998qj} is a powerful statement that has been used to explore nonperturbative aspects of quantum field theories for many years.
On the field theory side of the AdS/CFT, a conformal field theory is by definition a UV complete framework, in which a local quantum field theory exists at all energy scales.
This remains true for relevant or marginal deformations of CFTs which preserve the existence of a UV fixed point.
Then, a natural question arises: can the AdS/CFT be extended to a correspondence between nearly AdS and nearly CFT with an irrelevant deformation of CFTs?
In the context of AdS$_{3}$/CFT$_{2}$ correspondence, McGough, Mezei, and Verlinde have recently proposed an intriguing extension of the AdS/CFT \cite{McGough:2016lol}, which is based on the so-called integrable $T\bar{T}$ deformation of CFT \cite{Smirnov:2016lqw,Cavaglia:2016oda}.
On the bulk side, the boundary lies not at asymptotic infinity, but instead is located at a finite radial position.
The dual quantum field theory is no longer conformal, but is described by a QFT deformed by the remarkable $T\bar{T}$ operator of Zamolodchikov \cite{Zamolodchikov:2004ce}.
This deformed integrable quantum field theory can be regarded as an effective field theory with a finite UV cutoff.
The bulk side of this proposed duality offers an interesting viewpoint: AdS/CFT with a finite boundary.
Moving the boundary inward could shed light on the question of the emergence of bulk locality.
The notion of introducing a cutoff boundary surface in this context has been known in earlier work, e.g.~\cite{Balasubramanian:1999jd,deBoer:1999tgo}, and also in relation to the fluid-gravity correspondence \cite{Brattan:2011my}.
In particular, \cite{Heemskerk:2010hk,Faulkner:2010jy} show that such cutoffs are dual to some deformation of the original CFT.

The $T\bar{T}$ deformation of two dimensional CFTs provides an exactly solvable model of quantum field theory with a UV cutoff \cite{Smirnov:2016lqw,Cavaglia:2016oda}.
Any CFT can be deformed by this operator, defining a one-parameter family of theories labelled by a dimensionful deformation parameter $\mu$.
Here we take $\mu$ to be positive, and the deformation is written as
\begin{equation}
S_{\text{QFT}}=S_{\text{CFT}}+ \mu \int d^{2}x \, T\bar{T},
\label{eq:deformedaction}
\end{equation}
where $T\bar{T}$ denotes the composite irrelevant operator given by the product of the left- and right-moving components of the stress tensor \cite{Zamolodchikov:2004ce}.
By finite $\mu$ we mean that there is a one parameter family of theories defined by $dS^{(\mu)}_{\text{QFT}} /d\mu = \int d^{2}x (T\bar{T})_{\mu}$, where the $(T\bar{T})_{\mu}$ emphasizes that in this equation we have to use the stress tensor defined through $S^{(\mu)}_{\text{QFT}}$.
The system is exactly solvable under this deformation \eqref{eq:deformedaction}, in the sense that the deformed theory also possesses an infinite set of conserved charges and allows one to exactly compute interesting physical quantities.
The deformation also admits equivalent descriptions in terms of a coupling to two-dimensional topological gravity and in terms of a random geometry \cite{Dubovsky:2012wk,Cardy:2018sdv}.

Conversely, the $T\bar{T}$ deformation opens a new window to study the AdS/CFT correspondence itself.
It is a double-trace deformation, and could change the boundary condition of the AdS gravity.
For a $T\bar{T}$ deformed holographic CFT,
%McGough, Mezei, and Verlinde
a deformed AdS/CFT has been proposed, in which the dual AdS$_{3}$ gravity should be defined in a finite region, with the asymptotic boundary being at a finite radial position \cite{McGough:2016lol}.
More precisely if a CFT has a gravity dual,
%i.e. it is a holographic CFT,
then the deformed theory is dual to the original gravitational theory with the new boundary at $r = r_{c}$.
The relation between the deformation parameter $\mu$ and the finite radial position $r_{c}$ is given by
\begin{equation}
\mu = \frac{16\pi G_{N}}{r_{c}^{2}} = \frac{24\pi}{c}\frac{1}{r_{c}^{2}},
\label{eq:newboundary}
\end{equation}
where we set the AdS radius to be $1$, and $c$ is the central charge of the original CFT.
This new correspondence has been checked from various points of view, and
more on the holographic interpretation of the $T\bar{T}$ deformation has also been studied, see e.g.~\cite{Giveon:2017nie,Dubovsky:2017cnj,Asrat:2017tzd,Giribet:2017imm,Kraus:2018xrn,Chakraborty:2018kpr,Hartman:2018tkw,Caputa:2019pam}.

In a holographic CFT, the entanglement entropy could be captured by the area of the minimal surface extended into the dual AdS bulk geometry via the Ryu-Takayanagi formula \cite{Ryu:2006bv,Ryu:2006ef}, whose gravitational derivation was given in \cite{Lewkowycz:2013nqa}, and entanglement entropy in general captures a great deal of the quantum structure of strongly interacting systems.
%Also, the holographic entanglement entropy can probe an information of quantum gravity.
As we have seen, according to \cite{McGough:2016lol} the $T\bar{T}$ deformed geometry on the gravity side is a cutoff AdS.
Thus, the holographic entanglement entropy may be directly affected by the deformation, and in fact it is found that entanglement entropies in the $T\bar{T}$ deformed CFTs generically have corrections due to the deformation, see e.g.~\cite{Donnelly:2018bef,Chen:2018eqk,Banerjee:2019ewu}.
In this paper, we focus on the two-interval holographic entanglement entropy \cite{Calabrese:2009ez,Calabrese:2010he,Headrick:2010zt,Hartman:2013mia} and the holographic entanglement of purification \cite{Takayanagi:2017knl,Nguyen:2017yqw} in cutoff AdS.
By considering the new duality proposed in \cite{McGough:2016lol}, assuming that the RT formula still holds, we study the holographic entanglements in the $T\bar{T}$ deformed CFTs.
In particular, we consider phase transitions of holographic entanglements in the cutoff AdS and investigate its implication in the AdS/CFT correspondence.
Our results imply that the subsystems effectively become smaller under the $T\bar{T}$ deformation, that is, the degrees of freedom in the subsystems are reduced.
Our results are consistent with \cite{Park:2018snf}, in which the entanglement entropy of a boosted system is computed and the Lorentz-contracted subsystem is regarded as a $T\bar{T}$ deformed system.
For the entanglement of purification, we also find that in the $T\bar{T}$ deformed CFTs a phase transition can occur for fixed subsystems.

The rest of the paper is organized as follows.
In Sec.~\ref{Sec:twointerval}, we consider two-interval holographic entanglement entropy and its phase transition. In the cutoff AdS, we will find that the transition point is monotonically decreasing, which means that the subsystem on the boundary effectively becomes smaller.
In Sec.~\ref{sec:eop}, we also investigate the effect of the $T\bar{T}$ deformation on the entanglement wedge cross section, which is conjectured to be a gravity dual of the entanglement of purification in holographic CFTs \cite{Takayanagi:2017knl}.
In Sec.~\ref{sec:conclusion}, we present our conclusions and discussion.

%%%%%%%%%%%%%%%%%% section 2 %%%%%%%%%%%%%%%%%%%%%%%%%%%%%%%%%
\section{Two-interval entanglement entropy}
\label{Sec:twointerval}

In this section we discuss two-interval holographic entanglement entropy and its transition in cutoff AdS.
For this purpose, we need to compute the lengths of geodesics in cutoff AdS and to compare them.
By solving for the transition point numerically, we will find that it decreases monotonically with the deformation.
On the field theory side, this result can be interpreted as indicating that the degrees of freedom of the subsystems are decreasing.

%%%%%%%%%%%%%%%%%%%%%%%%%%%%%%%%%%%%%%%%
\subsection{Geodesic in cutoff AdS}
\label{sec:geodesic}

For later use, let us begin with the computation of a geodesic in $\mathrm{AdS_{3}}$.
The pure $\mathrm{AdS_{3}}$ metric in Poincar\'e coordinates is
\begin{equation}
ds^{2}=\frac{-dt^{2}+dx^{2}+dz^{2}}{z^{2}},
\end{equation}
where the AdS radius is set to $1$.
%For simplicity, we set $R=1$ in what follows.
For a static curve in $\mathrm{AdS_{3}}$ at a canonical time slice, the induced metric on it is given by
\begin{align}
ds_{static}^{2} & =\frac{1}{z(x)^{2}}(dx^{2}+z'(x)^{2}dx^{2})\nonumber \\
 & =\frac{1+z'(x)^{2}}{z(x)^{2}}dx^{2},
\end{align}
where we have employed the static gauge.
The Euler-Lagrange equation for the length functional of the curve
\begin{equation}
L[z]=\int\frac{\sqrt{1+z'(x)^{2}}}{z(x)}dx
\end{equation}
yields the equation of motion for $z(x)$. Solving it with the boundary conditions
\begin{equation}
z'(0)=0
\end{equation}
and
\begin{equation}
z(0)=0\;\mathrm{or}\;z(a)=0
\end{equation}
yields semicircular solutions of radius $a$.

%FIGURE
\begin{figure}[tbp]
\centering
\includegraphics[width=6cm, trim=5cm 4cm 5cm 2cm]{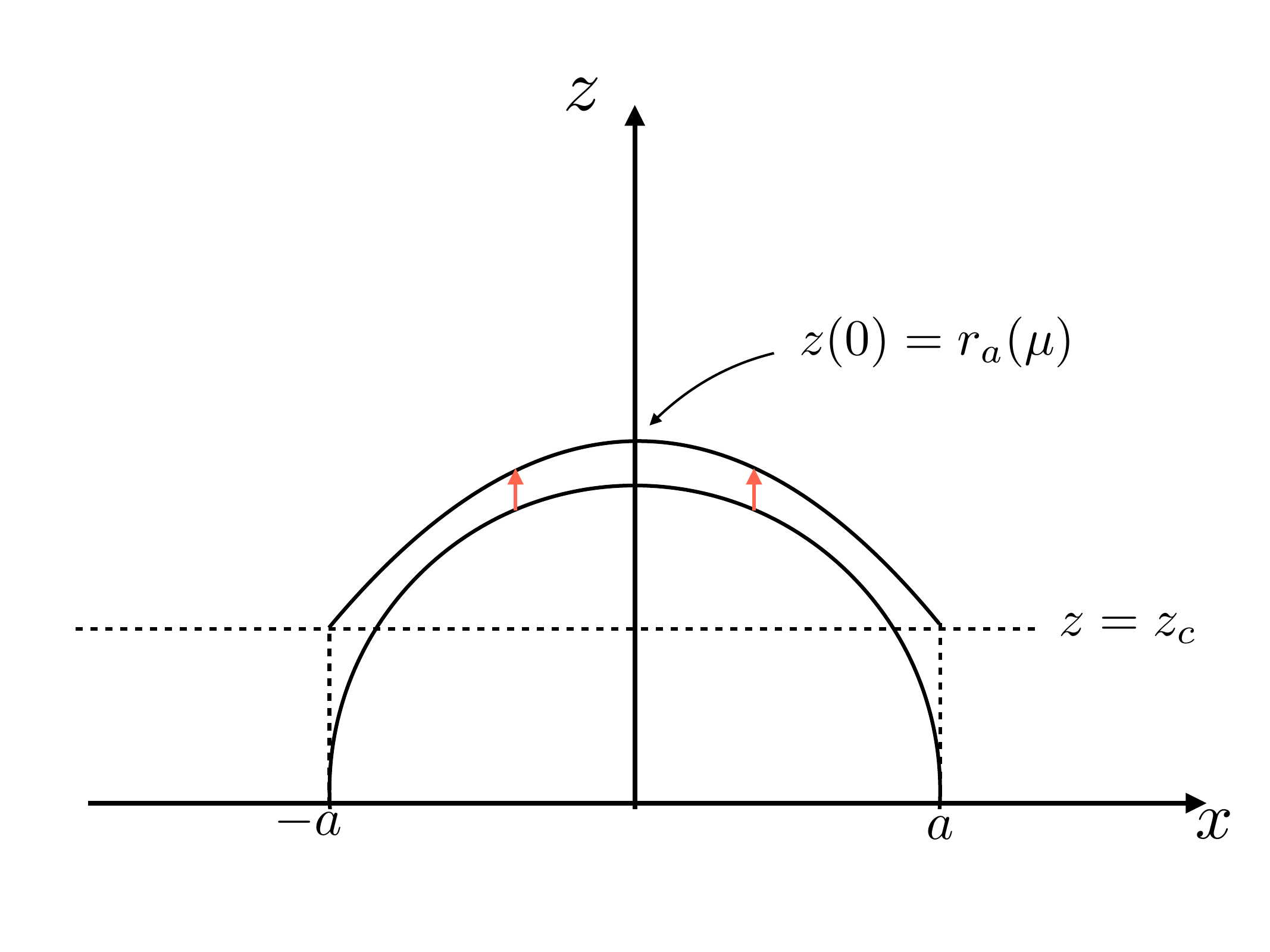}
\caption{A deformed geodesic in the $T\bar{T}$ deformed geometry. }
\label{fig:deformed}
\end{figure}

Let us move on to the $T\bar{T}$ deformed geometry, namely cutoff AdS.
We introduce a finite radial cutoff \cite{McGough:2016lol}
\begin{equation}
r_{c}^{2}=\frac{24\pi}{\mu c}.
\end{equation}
In $z$ coordinate, the radial cutoff is located at
\begin{equation}
z_{c}=\frac{1}{r_{c}}=\sqrt{\frac{\mu c}{24\pi}}.
\end{equation}
We now have $\mathrm{AdS_{3}}$ spacetime with a Dirichlet boundary surface at $z=z_{c}$:
\begin{equation}
ds^{2}=\frac{-dt^{2}+dx^{2}+dz^{2}}{z^{2}},\;z_{c}\leq z\leq\infty.
\end{equation}
Then, the length functional yields the same equation of motion as before, but we need to impose the boundary conditions
\begin{equation}
z'(0)=0
\end{equation}
and
\begin{equation}
z(a)=\sqrt{\frac{\mu c}{24\pi}} = z_{c}.
\end{equation}
A solution is given by
\begin{equation}
z(x)=\left(a^{2}+z_{c}^{2}-x^{2}\right)^{1/2},\;-a\leq x\leq a.\label{eq:dgeodesic}
\end{equation}
This solution is schematically drawn in Fig.~\ref{fig:deformed}, where we write the deformed radius as
\begin{equation}
z(0)=\left(a^{2}+z_{c}^{2}\right)^{1/2}=:r_{a}(\mu).
\end{equation}
In this kind of deformed geometry, the corrections to the holographic entanglement entropy have been calculated and appear to be consistent with those of the deformed CFT \cite{Donnelly:2018bef, Chen:2018eqk}.

%%%%%%%%%%%%%%%%%%%%%%%%%%%%%%%%%%%%%%%%

\subsection{Two-interval entanglement entropy in cutoff AdS}
\label{sec:twointerval}

\begin{figure}[tbp]
\begin{minipage}{0.5\hsize}
\begin{center}
\includegraphics[width=7cm,trim=0cm 3cm 0cm 2cm]{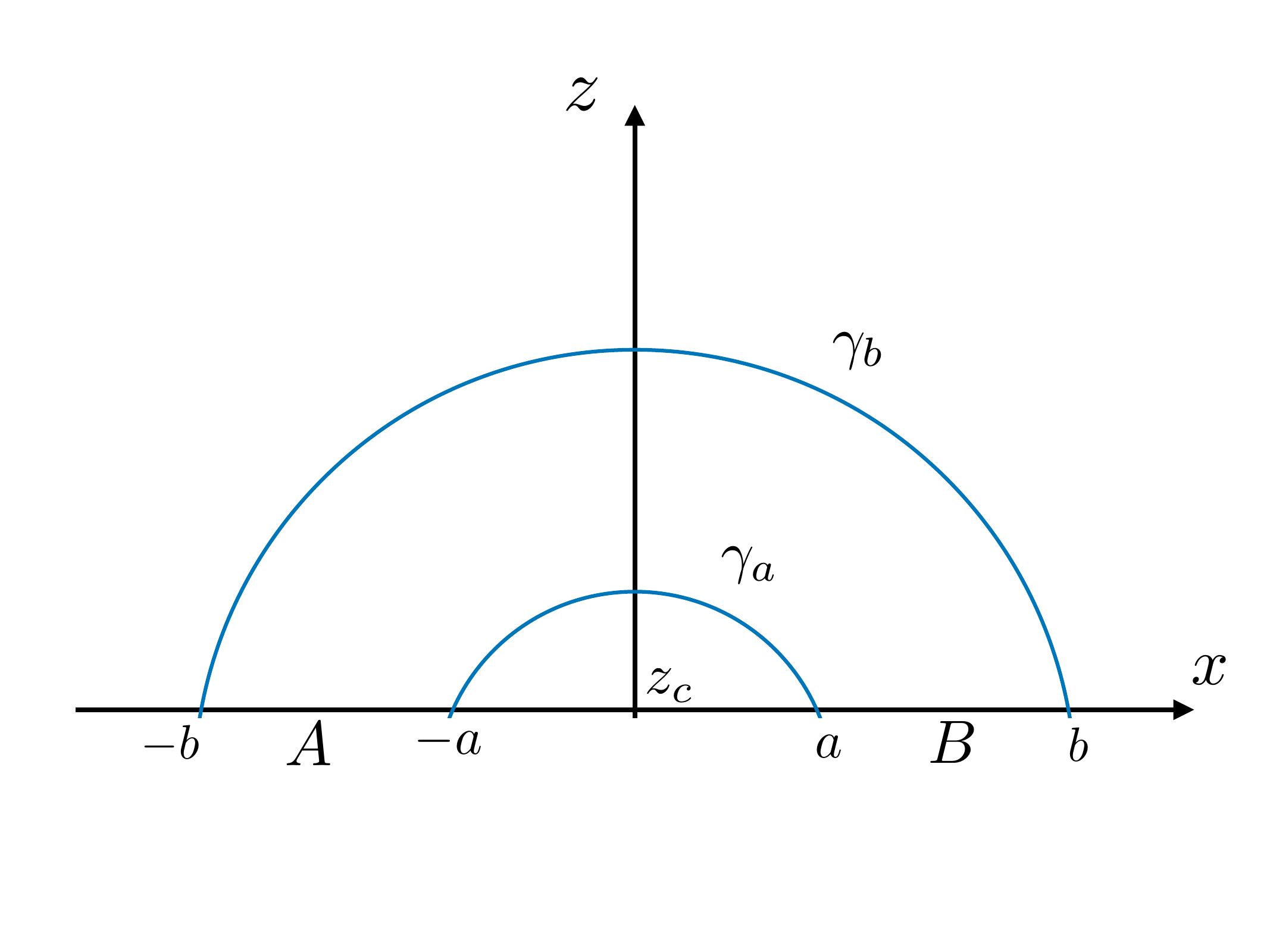}
\hspace{3.0cm} $(a)$
\end{center}
\end{minipage}
\begin{minipage}{0.5\hsize}
\begin{center}
\includegraphics[width=7cm,trim=0cm 3cm 0cm 2cm]{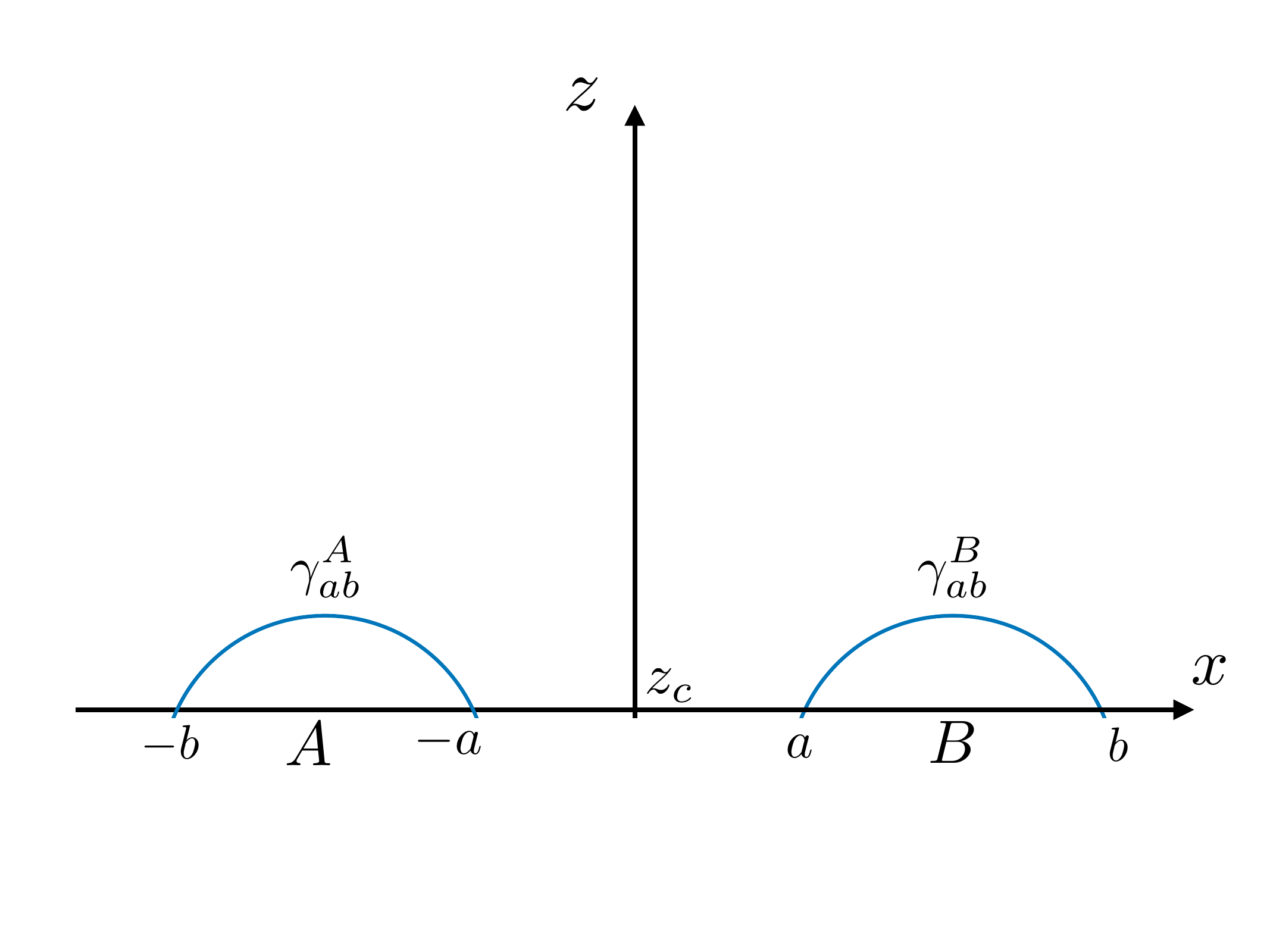}
\hspace{3.0cm} $(b)$
\end{center}
\end{minipage}
\caption{Two configurations of geodesics for two-interval holographic entanglement entropy. }
\label{fig:twointerval}
\end{figure}

Let us now consider two-interval entanglement entropy and its transition.
For simplicity, we focus on the symmetric configurations shown in Figs.~\ref{fig:twointerval} ($a$), ($b$).
There are two choices of geodesics ending on the endpoints of the subsystems $A$ and $B$.
For two-interval subsystems in this configuration, the holographic entanglement entropy is given by \cite{Headrick:2010zt,Hartman:2013mia}
%Calabrese:2009ez,Calabrese:2010he,
\begin{equation}
S_{AB} = \frac{1}{4G_N} \min \{ \lambda_{a}+\lambda_{b},~2\lambda_{ab}\},
\end{equation}
where $\la_{a},\, \la_{b},\, \la_{ab}$ are the lengths of $\gamma_{a},\, \gamma_{b},\, \gamma_{ab}$, respectively.
The configuration that gives the entropy is the one in which the geodesics have minimal total length, and a phase transition between the two generically occurs.
To see the transition, we compute the lengths of the geodesics in Fig.~\ref{fig:twointerval} ($a$) and Fig.~\ref{fig:twointerval} ($b$), and compare $\lambda_{a}+\lambda_{b}$ and $2\lambda_{ab}$.
When $\lambda_{a}+\lambda_{b}\leq2\lambda_{ab}$, the configuration in Fig.~\ref{fig:twointerval} ($a$) is realized, while when $\lambda_{a}+\lambda_{b}\geq2\lambda_{ab}$, that in Fig.~\ref{fig:twointerval} ($b$) is realized.
We now compute the lengths and compare them.
%FIGURE

We first consider the case of Fig.~\ref{fig:twointerval} ($a$).
The geodesic $\gamma_{a}$ is, as we obtained before, given by Eq.~(\ref{eq:dgeodesic})
\begin{equation}
z(x)=\left(a^{2}+z_{c}^{2}-x^{2}\right)^{1/2},\;-a\leq x\leq a.
\end{equation}
To compute its length in the cutoff AdS, we introduce polar coordinates:
\begin{align}
x & =\alpha\cos\theta,\\
z & =\alpha\sin\theta,
\end{align}
 where we have defined $\alpha^{2}=a^{2}+z_{c}^{2}$.
Then, the induced metric on the geodesic is
\begin{align}
ds^{2} & =\frac{\alpha^{2}}{z^{2}}d\theta^{2}\nonumber \\
 & =\frac{1}{\sin^{2}\theta}d\theta^{2}.
\end{align}
Using this, the length of the geodesic $\lambda_{a}$ is computed as
\begin{equation}
\lambda_{a}=\int ds=2\int_{\theta_{a}}^{\pi/2}\frac{d\theta}{\sin\theta},
\end{equation}
where $\theta_{a}$ is defined by the relations $\cos\theta_{a}=a/\alpha,\,\sin\theta_{a}=z_{c}/\alpha$.
Performing the integral, we find that
\begin{align}
\lambda_{a} & =\log\left(\frac{\alpha+a}{\alpha-a}\right)\nonumber \\
 & =\log\left(\frac{\sqrt{1+z_{c}^{2}/a^{2}}+1}{\sqrt{1+z_{c}^{2}/a^{2}}-1}\right).
\end{align}
 For $\lambda_{b}$, the calculation is exactly the same,
%\begin{equation}
%\lambda_{b}=\lambda_{a}|_{a\rightarrow b}.
%\end{equation}
 so we obtain
\begin{equation}
\lambda_{a}+\lambda_{b}=\log\left(\frac{\sqrt{1+z_{c}^{2}/a^{2}}+1}{\sqrt{1+z_{c}^{2}/a^{2}}-1}\right)+\log\left(\frac{\sqrt{1+z_{c}^{2}/b^{2}}+1}{\sqrt{1+z_{c}^{2}/b^{2}}-1}\right).
\end{equation}

For the case of Fig.~\ref{fig:twointerval} ($b$), $2\lambda_{ab}$ can also be calculated using these results.
In the $T\bar{T}$ deformed geometry, the solution of the geodesic $\gamma_{ab}^{B}$ is given by
\begin{equation}
z(x)=\sqrt{\left(\frac{b-a}{2}\right)^{2}+z_{c}^{2}-\left(x-\frac{a+b}{2}\right)^{2}}.
\end{equation}
We thus obtain
\begin{align}
2\lambda_{ab} & =2\lambda_{a}|_{a\rightarrow(b-a)/2}\nonumber \\
 & =2\log\left(\frac{\sqrt{1+z_{c}^{2}/\left(\frac{b-a}{2}\right)^{2}}+1}{\sqrt{1+z_{c}^{2}/\left(\frac{b-a}{2}\right)^{2}}-1}\right).
\end{align}

%%%%%%%%%%%%%%%%%%%%%%%%%%%%%%%%%%%%%%%%

\subsection{Transition between the two cases}
\label{sec:transition}

As we have seen, the holographic entanglement entropy of the two intervals is given
by $\lambda_{a}+\lambda_{b}$ or $2\lambda_{ab}$ as
\begin{equation}
S_{AB}=\frac{c}{6}\min\left\{ \lambda_{a}+\lambda_{b},\;2\lambda_{ab}\right\},
\end{equation}
where we have used the relation $c=3/(2G_{N})$ \cite{Brown:1986nw}. Which
of the two is smaller is determined by the difference $b-a$, or in other
words the ratio $a/b$. We fix $b$ and change $a$ from $0$ to $b$
to find the transition, i.e.~by solving $\lambda_{a}+\lambda_{b}=2\lambda_{ab}$
with respect to $a$, we find the transition point. When $\mu c \rightarrow 0$\footnote{Since we are considering classical gravity, we have to take the large-$c$ limit in order to compare with the bulk dual. This should be a 't Hooft-like limit, in which the combination $\mu c$ is kept finite \cite{Aharony:2018vux}. Under this limit, $\mu c$ is kept constant and the corrections of entanglements will be proportional to $c$. },
$\lambda_{a}+\lambda_{b}=2\lambda_{ab}$ becomes
\begin{equation}
2a\times2b=(b-a)^{2},
\end{equation}
 so when we change $a$, the transition occurs at
\begin{equation}
a_{0}=(3\pm2\sqrt{2})b.
\end{equation}
 Since we are considering $a\in[0,b]$,  the solution must be $a_{0}=(3-2\sqrt{2})b.$

%Figure
\begin{figure}[tbp]
\centering
\includegraphics[width=6.5cm, trim=5cm 4cm 5cm 2cm]{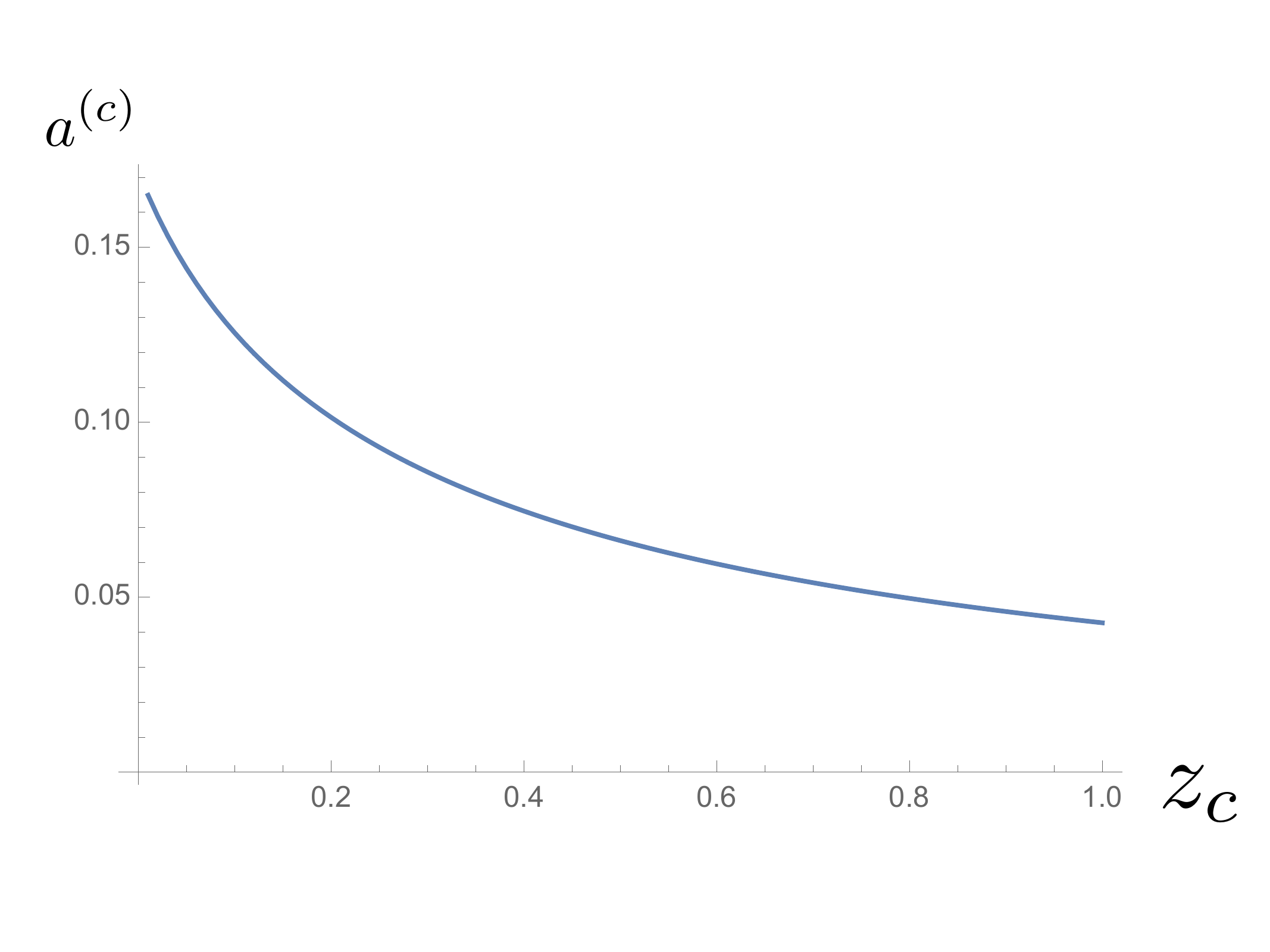}
\caption{The transition point. In this plot we set $b=1$. }
\label{fig:transition}
\end{figure}

Now, let us consider entanglement entropies with finite $\mu c$.
The equation we need to solve is
\begin{equation}
\left(\frac{\sqrt{1+z_{c}^{2}/\left(\frac{b-a}{2}\right)^{2}}+1}{\sqrt{1+z_{c}^{2}/\left(\frac{b-a}{2}\right)^{2}}-1}\right)^{2}=\frac{\sqrt{1+z_{c}^{2}/a^{2}}+1}{\sqrt{1+z_{c}^{2}/a^{2}}-1}\frac{\sqrt{1+z_{c}^{2}/b^{2}}+1}{\sqrt{1+z_{c}^{2}/b^{2}}-1}.
\end{equation}
This can be rewritten as
\begin{equation}
\left(\frac{b-a}{2}\right)^{2}\left(\sqrt{1+z_{c}^{2}\left(\frac{b-a}{2}\right)^{-2}}+1\right)^{2}=ab\left(\sqrt{1+\frac{z_{c}^{2}}{a^{2}}}+1\right)\left(\sqrt{1+\frac{z_{c}^{2}}{b^{2}}}+1\right).\label{eq:trans}
\end{equation}
 We can see an asymptotic form of the solution to Eq.~(\ref{eq:trans})
in the small $z_{c}$ region. When $z_{c}/a,\,z_{c}/b\ll1$, Eq.~(\ref{eq:trans})
can be expanded for small deformations and solved for $a$ perturbatively
to order $\mathcal{O}(z_{c}/b)$:
\begin{align}
\frac{a^{(c)}}{z_{c}} & =(3-2\sqrt{2})\frac{b}{z_{c}}-\frac{\sqrt{2}}{2}\frac{z_{c}}{b}\nonumber \\
 & =\frac{a_{0}}{z_{c}}-\frac{\sqrt{2}}{2}\frac{z_{c}}{b}.
\end{align}
We can also solve \eqref{eq:trans} for $a$ numerically, see Fig.~\ref{fig:transition}.
For a constant $b$, this result means that under the $T\bar{T}$ deformation two intervals need to get closer for the transition to occur.
Entanglement entropy generally measures correlations between subsystems, so our result indicates that the degrees of freedom in the subsystems are decreasing.
Thus we conclude that the subsystems effectively become smaller under the deformation.
If we take $\mu c \to 0$, as we have seen, the original CFT result is reproduced \cite{Headrick:2010zt,Hartman:2013mia}.
% which is consistent picture described in \cite{Park:2018snf}.
%Physical meaning
%Also, we can read off another fact. Even for large deformation, there is still a finite transition point. This is not so surprising and it reflects that AdS space is homogeneous.

%%%%%%%%%%%%%%%%%% section 3 %%%%%%%%%%%%%%%%%%%%%%%%%%%%%%%%%
\section{Holographic EoP in cutoff AdS}
\label{sec:eop}

\subsection{Entanglement wedge cross section in pure AdS}
\label{sec:eoppure}

\begin{figure}[tbp]
\centering
\includegraphics[width=6cm, trim=5cm 4cm 5cm 2cm]{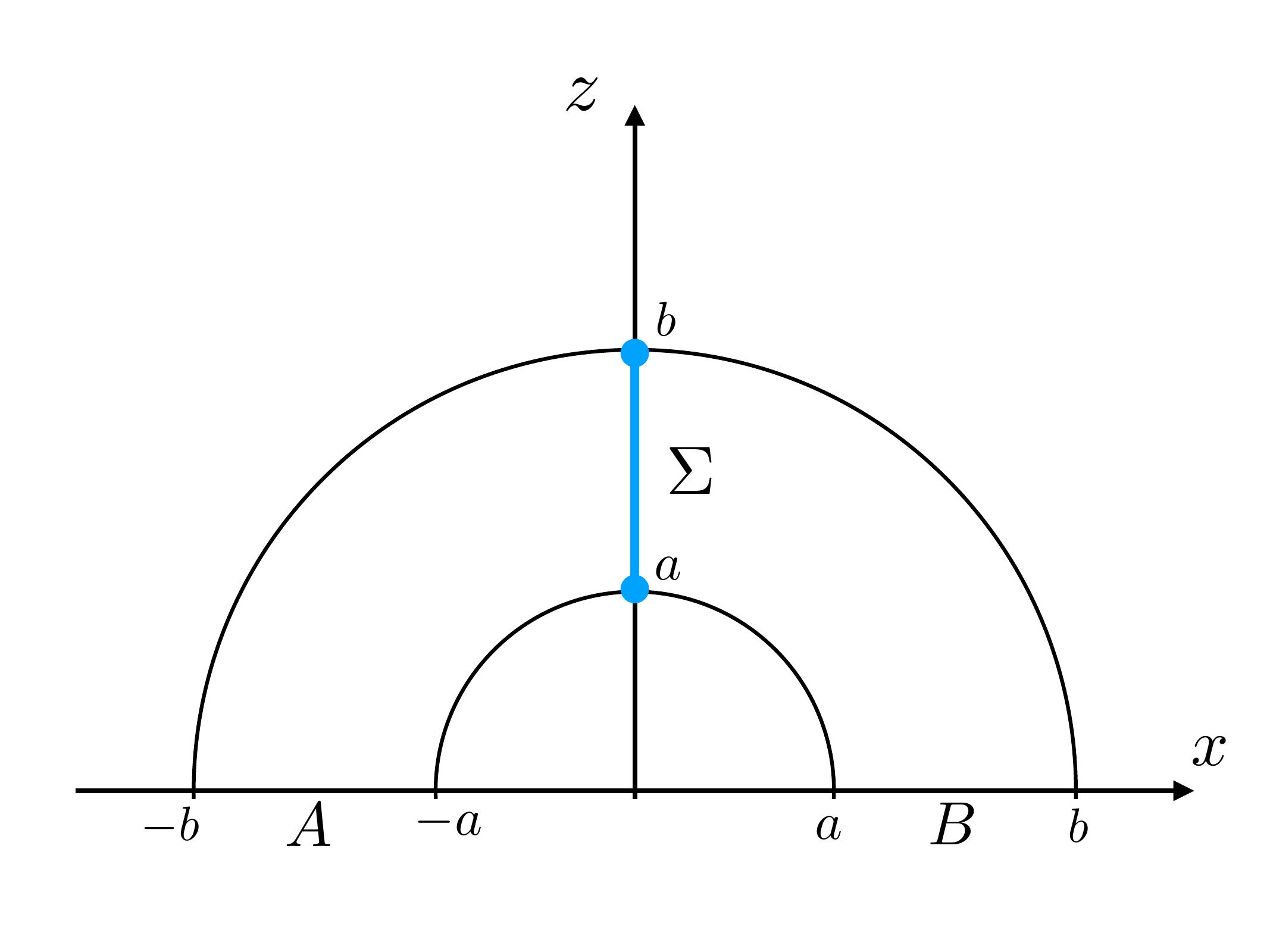}
\caption{Entanglement wedge cross section. }
\label{fig:crosssection}
\end{figure}

In this section, we discuss the entanglement wedge cross section, which is conjectured to be holographically dual to the entanglement of purification \cite{Terhal:2002riz} of the boundary state \cite{Takayanagi:2017knl,Nguyen:2017yqw}, in cutoff AdS both at zero temperature and at finite temperature.
The entanglement wedge cross section is also known to compute other measures of mixed state correlations, see e.g.~\cite{Umemoto:2018jpc,Tamaoka:2018ned,Kudler-Flam:2018qjo}.
Here again, for simplicity we consider a symmetric configuration described in Fig.~\ref{fig:crosssection}.
To have a connected entanglement wedge, we require $\la_{a}+\la_{b} < 2\la_{ab}$, where $\la$'s are the lengths of geodesics in the previous section.
Before turning to the $T\bar{T}$ deformed geometry, we first review the computation of the entanglement wedge cross section in pure AdS.
The length of $\Sigma$ in Fig.~\ref{fig:crosssection}
is given by
\begin{equation}
\Acal
%=\int ds
=\int_{a}^{b}\frac{dz}{z}=\log\frac{b}{a}.
\end{equation}
The entanglement wedge cross section is
\begin{equation}
E_{W}(A:B)=\frac{1}{4G_{N}}\log\frac{b}{a}=\frac{c}{6}\log\frac{b}{a},
\end{equation}
where we again have used the relation $c=3/(2G_{N})$.
%FIGURE

\begin{figure}
\centering
\includegraphics[width=6cm, trim=5cm 4cm 5cm 2cm]{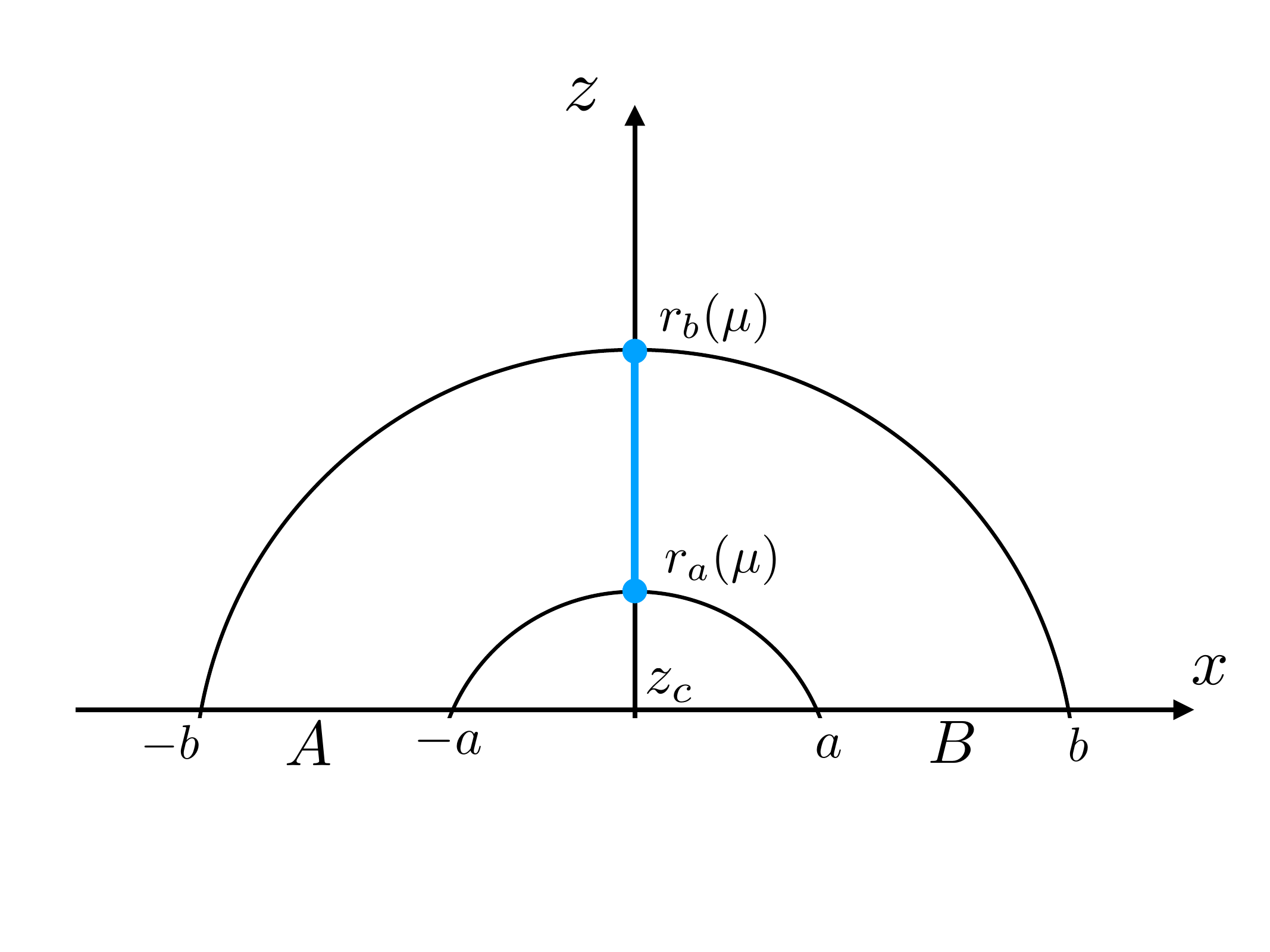}
\caption{Deformed entanglement wedge cross section. }
\label{fig:dcrosssection}
\end{figure}

In the $T\bar{T}$ deformed geometry, using the solution of the deformed
geodesics (\ref{eq:dgeodesic}), the entanglement wedge cross section
becomes (see Fig.~\ref{fig:dcrosssection})
\begin{align}
E_{W}^{\mu} & =\frac{c}{6}\log\frac{r_{b}(\mu)}{r_{a}(\mu)}\nonumber \\
 & =\frac{c}{6}\log\frac{b}{a}+\frac{c}{6}\log\left[\frac{1+z_{c}^{2}/b^{2}}{1+z_{c}^{2}/a^{2}}\right]^{1/2}.
\end{align}
%FIGURE
For $z_{c}^{2}/a^{2},\,z_{c}^{2}/b^{2}\ll1$, it can be expanded as
\begin{equation}
\left(1+\frac{z_{c}^{2}}{2b^{2}}\right)\left(1-\frac{z_{c}^{2}}{2a^{2}}\right)=1-\frac{\mu c}{48\pi}\frac{b^{2}-a^{2}}{a^{2}b^{2}}+\mathcal{O}(\mu^{2}),
\end{equation}
so for small enough deformations, the correction term of order $\Ocal(\mu^{1})$ is given by
\begin{align}
\Delta E_{W}^{\mu} & \simeq\frac{c}{6}\times\left(-\frac{\mu c}{48\pi}\frac{b^{2}-a^{2}}{a^{2}b^{2}}\right)\nonumber \\
 & =-\frac{\mu c^{2}}{288\pi}\frac{b^{2}-a^{2}}{a^{2}b^{2}}.
\end{align}
%where we have used the fact that
%\begin{equation}
%\log(1-x)=-\sum_{n=1}^{\infty}\frac{x^{n}}{n}.
%\end{equation}
This implies that in a $T\bar{T}$ deformed CFT the corresponding EoP will decrease even for a small deformation.
%The holographic entanglement entropy does not change under $T\bar{T}$ deformation of order $\mathcal{O}(\mu^{1})$, while $E_{W}^{\mu}$ does.
On the other hand, when one considers $z_{c}^{2}/a^{2},\,z_{c}^{2}/b^{2}\rightarrow\infty$,
\begin{equation}
E_{W}^{\mu}=\frac{c}{6}\log\left[\frac{b^{2}/z_{c}^{2}+1}{a^{2}/z_{c}^{2}+1}\right]^{1/2}\rightarrow0,
\end{equation}
which means that the dual CFT becomes trivial.

%%%%%%%%%%%%%%%%%%%%%%%%%%%%%%%%%%%%%%%%%%%%%%%%%

\subsection{Entanglement wedge cross section in BTZ black hole}
\label{sec:eopbtz}

\begin{figure}
\centering
\includegraphics[width=6cm, trim=5cm 4cm 5cm 2cm]{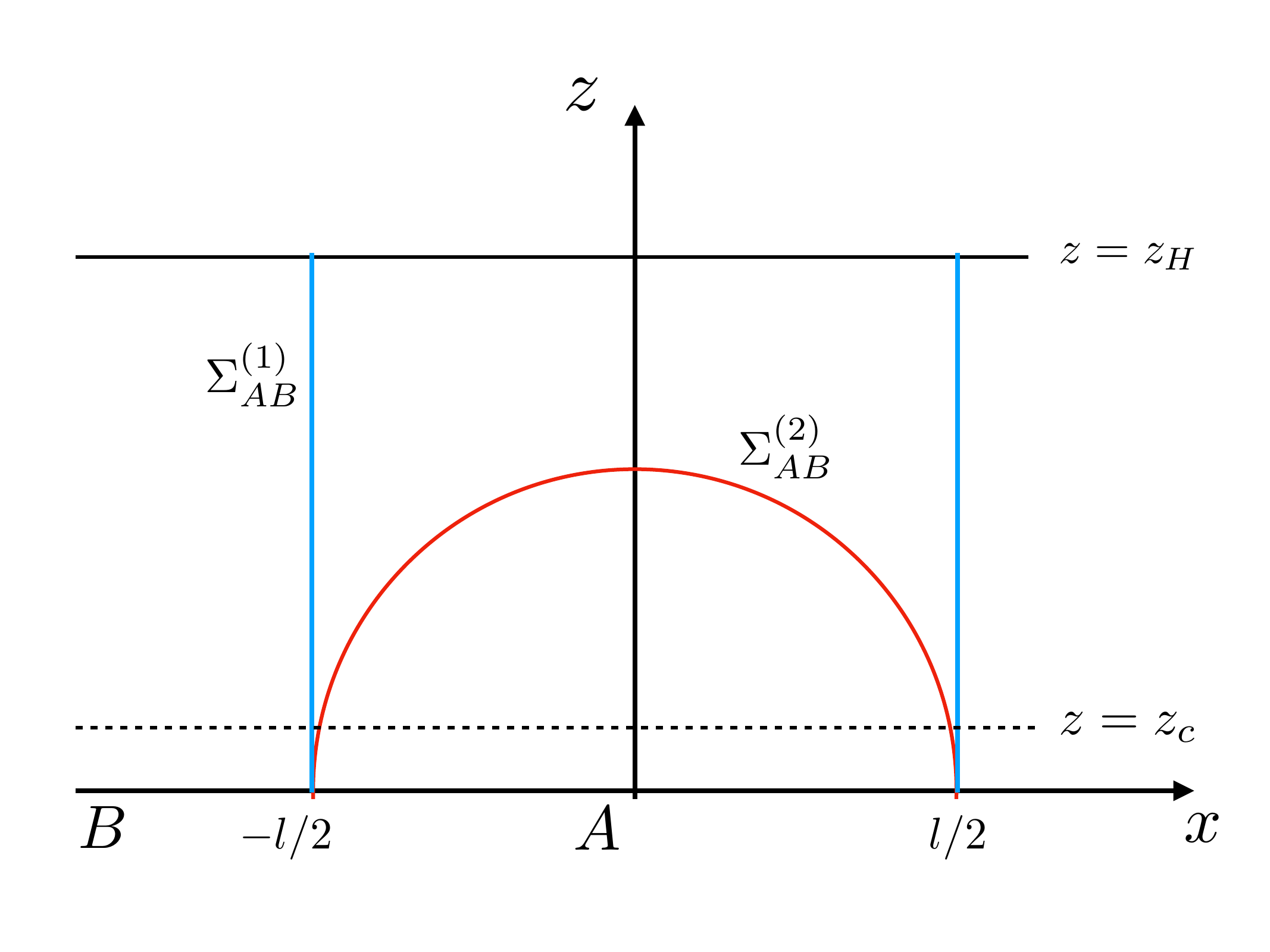}
\caption{Entanglement wedge cross section in BTZ black hole geometry}
\label{fig:btzcrosssection}
\end{figure}

We next consider a finite temperature state in a holographic CFT, which corresponds to a planar BTZ black hole.
The metric is given by
\begin{align}
ds^{2} & =\frac{1}{z^{2}}\left(-f(z)dt^{2}+\frac{dz^{2}}{f(z)}+dx^{2}\right),\\
f(z) & =1-\frac{z^{2}}{z_{H}^{2}},
\end{align}
where the location of the black hole horizon $z_{H}$ is related to the inverse temperature $\beta$ by $\beta=2\pi z_{H}$.
We define the subsystem $A$ to be the interval $-l/2 \leq x \leq l/2$ and $B$ to be its complement.
Then, we need to consider two cases $\Sigma_{AB}^{(1)}$ and $\Sigma_{AB}^{(2)}$ described in Fig.~\ref{fig:btzcrosssection}.
The entanglement wedge cross section is then
\begin{align}
E_{W}=\frac{c}{6}\min \{ 2 \Acal^{(1)},\, \Acal^{(2)} \},
\end{align}
where $\Acal^{(1)},\, \Acal^{(2)}$ are the lengths of $\Sigma_{AB}^{(1)}$ and $\Sigma_{AB}^{(2)}$.
Similarly to the analyses in the previous section, the favored entanglement wedge cross section will be the one in which the total length of the cross section is minimal.
%In field theory language, they correspond to entanglement of purification in thermal CFT.
%FIGURE

The length of $\Sigma_{AB}^{(1)}$ is given by
\begin{align}
\mathcal{A}^{(1)} & =\int_{z_{c}}^{z_{H}}\frac{1}{z\sqrt{f(z)}}dz\nonumber \\
% & =z_{H}\int_{z_{c}}^{z_{H}}\frac{dz}{z\sqrt{z_{H}^{2}-z^{2}}}
&=\log\left(\frac{z_{H}+\sqrt{z_{H}^{2}-z_{c}^{2}}}{z_{c}}\right),
\end{align}
 where $z_{c}=\sqrt{\mu c/24\pi}$.
 We thus obtain
\begin{align}
E_{W}^{(1)\,\mu}  =\frac{2\mathcal{A}^{(1)}}{4G_{N}}  =\frac{c}{3}\log\left(\frac{z_{H}+\sqrt{z_{H}^{2}-z_{c}^{2}}}{z_{c}}\right).
\end{align}
 In this geometry, $z_{c}$ must satisfy $0\leq z_{c}\leq z_{H}$, which means that
\begin{equation}
\mu c \leq\frac{6\beta^{2}}{\pi}.
\end{equation}
 This naively implies that the deformation in the thermal CFT is bounded by the temperature. Also, when $z_{c}\rightarrow z_{H}$, $\mathcal{A}^{(1)}\rightarrow0$.
If we consider $z_{c}/z_{H}\ll1$,
\begin{align}
\log\left(\frac{z_{H}+\sqrt{z_{H}^{2}-z_{c}^{2}}}{z_{c}}\right)
%& \simeq\log\left[\frac{z_{H}}{z_{c}}\left\{ 1+\left(1-\frac{1}{2}\left(\frac{z_{c}}{z_{H}}\right)^{2}\right)\right\} \right]\nonumber \\
 & \simeq \log\frac{2z_{H}}{z_{c}}+\log\left(1-\frac{z_{c}^{2}}{4z_{H}^{2}}\right)\nonumber \\
 & =\log\left(\frac{\beta}{\pi z_{c}}\right)-\frac{\pi}{24}\frac{\mu c}{\beta^{2}}.
\end{align}
 The first term is the ordinary contribution, in which $z_{c}$ simply
corresponds to a small cutoff.
So, the correction term of order $\mathcal{O}(\mu^{1})$ can be written as
\begin{equation}
\Delta E_{W}^{(1)\,\mu}=-\frac{\pi}{72}\frac{\mu c^{2}}{\beta^{2}}.
\end{equation}

We next consider the case of $\Sigma_{AB}^{(2)}$.
For the computation, we write the BTZ black hole metric in global coordinates:
\begin{equation}
ds^{2}=-\cosh^{2}\rho d\tau^{2}+d\rho^{2}+\sinh^{2}\rho d\theta^{2},
\end{equation}
 where
\begin{equation}
z=\frac{z_{H}}{\cosh\rho},
\end{equation}
and so
\begin{equation}
\cosh\rho_{\infty}=\frac{z_{H}}{z_{c}}.
\end{equation}
Then, the length of $\Sigma_{AB}^{(2)}$ is given by
\begin{align}
\cosh\mathcal{A}^{(2)} & =1+2\cosh^{2}\rho_{\infty}\sinh^{2}\frac{\pi l}{\beta}\nonumber \\
 & =1+2\left(\frac{z_{H}}{z_{c}}\right)^{2}\sinh^{2}\frac{\pi l}{\beta}.
\end{align}
 We then find that
\begin{equation}
\mathcal{A}^{(2)}=\log\left[1+2\left(\frac{z_{H}}{z_{c}}\right)^{2}\sinh^{2}\frac{\pi l}{\beta}\left(1+\sqrt{1+\left(\frac{z_{c}}{z_{H}}\right)^{2}\left(\sinh\frac{\pi l}{\beta}\right)^{-2}}\right)\right].
\end{equation}
For $z_{c}/z_{H}\ll1$, to order $\mathcal{O}(z_{c}/z_{H})$,
\begin{align}
\mathcal{A}^{(2)} & \simeq\log\left(2+4\left(\frac{z_{H}}{z_{c}}\right)^{2}\sinh^{2}\frac{\pi l}{\beta}\right)\nonumber \\
 & \simeq2\log\left(\frac{\beta}{\pi z_{c}}\sinh\frac{\pi l}{\beta}\right),
\end{align}
which is the well-known result.
Taking the second order into account, we obtain
\begin{equation}
\mathcal{A}^{(2)}\simeq\log\left[4\left(\frac{z_{H}}{z_{c}}\right)^{2}\sinh^{2}\frac{\pi l}{\beta}+2-\frac{1}{4}\left(\frac{z_{c}}{z_{H}}\right)^{2}\left(\sinh\frac{\pi l}{\beta}\right)^{-2}\right],
\end{equation}
 where
\begin{equation}
\left(\frac{z_{c}}{z_{H}}\right)^{2}=\frac{\pi}{6}\frac{\mu c}{\beta^{2}}.
\end{equation}

 When $z_{c}\rightarrow z_{H}$,
\begin{equation}
\mathcal{A}^{(2)}=\log\left[1+2\sinh^{2}\frac{\pi l}{\beta}\left(1+\sqrt{1+\left(\sinh\frac{\pi l}{\beta}\right)^{-2}}\right)\right]\neq0,
\end{equation}
 which means that a phase transition must occur, $\Sigma_{AB}^{(2)}\rightarrow\Sigma_{AB}^{(1)}$.
The transition occurs at the point $\mathcal{A}^{(2)}=2\mathcal{A}^{(1)}$, which gives
\begin{equation}
z_{c}^{(c)}=z_{H}\sqrt{1-\sinh^{2}\frac{\pi l}{\beta}}.
\label{eq:transpoint}
\end{equation}
 Thus, the entanglement wedge cross section is again given by
\begin{align}
E_{W} & =\frac{c}{6}\min\left\{ 2\mathcal{A}^{(1)},\;\mathcal{A}^{(2)}\right\}.
%0 & \leq z_{c}\leq z_{c}^{(c)}\;\Longrightarrow\;\Sigma_{AB}^{(2)},\\
%z_{c}^{(c)} & \leq z_{c}\leq z_{H}\;\Longrightarrow\;\Sigma_{AB}^{(1)}.
\end{align}
When $0  \leq z_{c}\leq z_{c}^{(c)}$ (which requires $\sinh(\pi l/\beta)<1$), $\Sigma_{AB}^{(2)}$ is realized, and   when $z_{c}^{(c)}  \leq z_{c}\leq z_{H}$, $\Sigma_{AB}^{(1)}$ is realized.
One comment regarding this is that we do not change the size of the subsystem, but only the deformation parameter.
This means that when we consider the $T\bar{T}$ deformation, the transition between the two entanglement wedge cross sections can occur for fixed subsystems.

 Let us translate them into CFT language.
 On the CFT side, the variables are interpreted through
\begin{align}
z_{c} =\sqrt{\frac{\mu c}{24\pi}}, ~ \beta =2\pi z_{H}.
\end{align}
 So the transition point \eqref{eq:transpoint} is rewritten in terms of the CFT deformation parameter $\mu$ and the inverse temperature $\beta$ by
\begin{align}
%\frac{\mu^{(c)}c}{24\pi} & =(\frac{\beta}{2\pi})^{2}\left(1-\sinh^{2}\frac{\pi l}{\beta}\right)\nonumber \\
%\Longleftrightarrow\,
\mu^{(c)} c & =\frac{6\beta^{2}}{\pi}\left(1-\sinh^{2}\frac{\pi l}{\beta}\right).
\end{align}
 Recall that from $z_{c}\leq z_{H}$, the deformation parameter $\mu$ is bounded by
\begin{align}
%\frac{\mu\pi c}{6} & \leq\frac{\beta^{2}}{4\pi^{2}}\nonumber \\
%\Longleftrightarrow\,
\mu c & \leq\frac{6\beta^{2}}{\pi}=:\mu_{H} c.
\end{align}
 Therefore, on the CFT side we find that
\begin{align}
0 & \leq\mu c \leq\mu^{(c)} c \;\Longrightarrow\;E_{P}^{(2)\,\mu},\\
\mu^{(c)} c & \leq\mu c \leq\mu_{H} c \;\Longrightarrow\;E_{P}^{(1)\,\mu}.
\end{align}
This implies that in a thermal CFT, even if one does not change the size of the subsystems, the $T\bar{T}$ deformation will induce a transition of the entanglement of purification.
To analyze this directly from the field theory side is challenging, since we do not know how to compute the entanglement of purification in a general QFT (although see \cite{Caputa:2018xuf} for a proposal in holographic CFTs).
Our result provides a holographic prediction for the effect of the integrable $T\bar{T}$ deformation on the entanglement of purification in deformed CFTs.

%%%%%%%%%%%%%%%%%% section 4 %%%%%%%%%%%%%%%%%%%%%%%%%%%%%%%%%
\section{Conclusion and Discussion}
\label{sec:conclusion}

%Let us first summarize the paper.
In this paper, we have carried out calculations of two-interval holographic entanglement entropy and holographic entanglement of purification in cutoff AdS, and also investigated the effect of the $T\bar{T}$ deformation on their phase transitions. In Sec.~\ref{Sec:twointerval}, we have studied the two-interval entanglement entropy in the cutoff AdS.
Fig.~\ref{fig:transition} shows that the subsystem becomes smaller under the $T\bar{T}$ deformation.
This is easy to understand on the gravity side, since the metric becomes small at larger $z$ and in the cutoff AdS the subsystems move into the bulk.
This is consistent with the viewpoint of a boosted CFT and its entanglement entropy \cite{Park:2018snf}, and it means that under the $T\bar{T}$ deformation the subsystems in the field theory effectively shrink, or in other words, the degrees of freedom in the subsystems are effectively reduced.
A natural further study is to check the behavior of the transition on the field theory side, and to evaluate to what extent the conjecture \cite{McGough:2016lol} is valid and how much the Ryu-Takayanagi formula can be assumed\footnote{While this paper was in preparation, a paper on the verification of the RT formula for $T\bar{T}$ deformed CFTs appeared \cite{Murdia:2019fax}. The authors of \cite{Murdia:2019fax} claim that if there exists a holographic duality between Einstein gravity in the bulk and a quantum field theory on the boundary such that the two are related by a GKPW-like relation, then the RT formula will hold. }.
This is difficult because the deformed quantum field theory is no longer conformal.
What one can do is either a fully nonperturbative analysis or a perturbative expansion around the original CFT.
Another question is the generalization to the multi-interval case.
When one considers a multi-interval case, it would be interesting to study holographic mutual information and the effect of the $T\bar{T}$ deformation on monogamy relations \cite{Hayden:2011ag}.

In Sec.~\ref{sec:eop}, we studied the effect of the $T\bar{T}$ deformation on the entanglement wedge cross section, which is conjectured to be dual to the entanglement of purification in the boundary field theory.
Concerning this, it would be interesting to consider wormholes, or eternal black holes in the $T\bar{T}$ deformed geometry.
They are originally constructed from the BTZ black hole geometry \cite{Maldacena:2001kr,Maldacena:2004rf}, and have been studied extensively.
%,Maldacena:2018gjk
The cutoff AdS, as we have seen, influences the entanglement entropy.
Generically the $T\bar{T}$ deformation therefore also affects thermofield double states, so that it may produce a deformed wormhole, on whose boundary the deformed CFTs live.

We found that two-interval holographic entanglement entropy in cutoff AdS does receive a correction due to the deformation, which implies that on the field theory side the degrees of freedom in a subregion are decreasing under the $T\bar{T}$ deformation.
Also, we investigated the entanglement wedge cross section in cutoff AdS both at zero temperature and at finite temperature.
The computation of the entanglement of purification in generic QFTs is difficult in general, so our results provide holographic predictions.
More study on holographic entanglements in this direction will play an important role in the understanding of holography beyond AdS/CFT.

\subsection*{Acknowledgements}

We would like to thank Naotaka Kubo, Kotaro Tamaoka and Koji Umemoto for valuable discussions.
The author thanks the Yukawa Institute for Theoretical Physics at Kyoto University, where this work was initiated during the workshop YITP-T-18-04 on ``Strings and Fields 2018.''
Discussions during the YITP Atoms program were also valuable in completing this work.
This work was supported by RIKEN Junior Research Associate Program.

%\appendix
%%%%%%%%%%%%%%%%%% section A %%%%%%%%%%%%%%%%%%%%%%%%%%%%%%%%
%\section{Definition and notation}

%%%%%%%%%%%%%%%%%% section B %%%%%%%%%%%%%%%%%%%%%%%%%%%%%%%%
%\section{Calculation detail}

%\clearpage
%%%%%%%%%%%%%%%%%% References %%%%%%%%%%%%%%%%%%%%%%%%%%%%%%%
%\input{Sections/Common/endbib}
%\bibliographystyle{Sections/Common/utphys}
%%\nocite{*}
%\bibliography{Sections/Common/Ref}
\providecommand{\href}[2]{#2}\begingroup\raggedright\endgroup

\end{document}